\documentclass[%
 aip,
 apl,%
 amsmath,amssymb,
 reprint,%
]{revtex4-1}

\usepackage[pdftex]{graphicx}
\usepackage{dcolumn}
\usepackage{bm}

\begin{document}

\preprint{AIP/123-QED}

\title[Sample title]{Brillouin lasing in coupled silica toroid microcavities}

\author{Yoshihiro Honda}
\author{Wataru Yoshiki}
\author{Tomohiro Tetsumoto}
\author{Shun Fujii}
\affiliation{Department of Electronics and Electrical Engineering, Faculty of Science and Technology, Keio University 3-14-1, Hiyoshi, Kohoku-ku, Yokohama 223-8522, Japan}
\author{Kentaro Furusawa}
\author{Norihiko Sekine}
\affiliation{Advanced ICT Research Institute, National Institute of Information and Communications Technology, 4-2-1, Nukuikitamachi, Koganei City, Tokyo 184-8795, Japan}
\author{Takasumi Tanabe}
 \email{takasumi@elec.keio.ac.jp}
\affiliation{Department of Electronics and Electrical Engineering, Faculty of Science and Technology, Keio University 3-14-1, Hiyoshi, Kohoku-ku, Yokohama 223-8522, Japan}

\date{\today}

\begin{abstract}
We demonstrate stimulated Brillouin scattering (SBS) lasing in a strongly coupled microcavity system.  By coupling two silica toroid microcavities, we achieve large mode splitting of 11~GHz, whose frequency separation matches the Brillouin frequency shift of silica.  The SBS light is resonantly amplified by pumping at the higher frequency side of the supermode splitting resonance. Since the mode splitting is adjusted by changing the gap distance between the two cavities, our system does not require precise control of a mm-sized cavity diameter to match the free-spectral spacing with the Brillouin frequency shift.  It also allows us to use a small cavity, hence our system has the potential to achieve the lasing threshold at a very low power.
\end{abstract}

\maketitle



The Brillouin scattering of light is a nonlinear process in which optical waves interact with an acoustic wave. When light is pumped, coherent light is generated due to backscattering via electrostriction caused by an acoustic wave.  The backward scattering light is Stokes light that has experienced a Doppler downshift and whose frequency is well defined due to the narrow bandwidth of the Brillouin gain.  This property has led to the Brillouin scattering in optical fibers and waveguides being well studied and various applications such as light storage \cite{Zhu1748}, slow light generation~\cite{PhysRevLett.94.153902,Agha:09}, high coherence lasers \cite{Smith:91} and microwave synthesizers \cite{Geng:08} have been demonstrated.

Since Brillouin scattering is a coherent process, Stokes light can be amplified with a stimulated process when it resonates with an optical mode.  Then stimulated Brillouin scattering (SBS) occurs.  Recently, SBS has been reported in whispering gallery mode (WGM) microcavities such as a crystalline resonator \cite{PhysRevLett.102.043902,doi:10.1063/1.4903516}, a microsphere \cite{PhysRevLett.102.113601,Bahl2011,Guo:15,Guo:15low}, a disk resonator \cite{Loh:15}, a rod resonator \cite{1367-2630-18-4-045001}, a microbubble/bottle resonator \cite{Lu:16,Asano:16} and a wedge resonator \cite{Lee2012,Li:12}. Because WGM microcavities have a high quality factor ($Q$) and a small mode volume \cite{Vahala2003}, we are able to generate SBS at a low input power whose SBS threshold power is proportional to the mode volume and inversely proportional to the square of the $Q$ factor.  Brillouin lasing using a silica wedge resonator has also been demonstrated as a low-phase-noise oscillator \cite{Li2013}.

However, the use of a single WGM microcavity for Brillouin lasing has certain limitations. To achieve Brillouin lasing, the frequency separation of two optical cavity modes (i.e., one for the pump mode and the other for the Stokes mode) must exactly match the Brillouin frequency shift.  Two different approaches have been demonstrated that enable Brillouin lasing using such a microcavity.

The first approach uses a large cavity with a precisely controlled free spectral range (FSR).  The designed FSR has exactly the same frequency as the Brillouin frequency shift \cite{Li:12,Loh:15}, to enable resonances at the pump and Stokes lights.   Since the FSR of a microcavity is determined by the size, a cavity must be fabricated with high precision, which is usually not an easy task.  This approach is possible for a proof-of-principle experiment but is unsuitable for applications because trial-and-error is required during fabrication.  In addition, the Brillouin frequency shift is several tens of GHz for silica glass, and so the fabricated cavity diameter must be close to a millimeter to obtain a GHz-FSR.  Therefore, the SBS threshold increases due to the large size of the cavity.

The second method employs different transverse modes in a microcavity for the Stokes and the pump \cite{PhysRevLett.102.043902,doi:10.1063/1.4903516,PhysRevLett.102.113601,Bahl2011,Guo:15,Guo:15low}.  WGM microcavities are often multimode, and so the researcher can search for a pair of modes that exactly matches the Brillouin frequency shift.  However, a deterministic approach is quite difficult, and usually researchers rely on trial-and-error until they find a pair of modes with exact frequency separation.

For both approaches, it is necessary to control the cavity shape and size precisely in order to demonstrate a deterministic approach.  Although trials for designing a WGM microcavity with the desired shape and size have just begun \cite{Lee2012,Nakagawa:16}, it remains a challenging task.

In this letter, we experimentally demonstrate Brillouin lasing in coupled microcavities to overcome the above problems.  Little attention has been paid to Brillouin lasing using coupled microcavities, and although the approach has been suggested using a double disk microcavity, it has yet to be demonstrated experimentally \cite{Espinel2017}.

The coupled cavity system forms supermodes that consist of symmetric and anti-symmetric modes, resulting in spectrum splitting in the resonant mode. Because the supermode splitting is dependent on the coupling strength between two modes, we can adjust the mode splitting frequency with the Brillouin frequency shift to enable resonant excitation of the pump and the probe.  We use silica toroid microcavities because they can be fabricated on a chip and have ultra-high $Q$s and a small mode volume \cite{Vahala2003}. Coupled silica toroid microcavities have already been used to amplify optomechanical oscillation \cite{PhysRevLett.104.083901} and all-optical tunable buffering \cite{Yoshiki2017}, but have not been employed for Brillouin lasing.  The challenge and the requirement as regards realizing Brillouin lasing in coupled microcavities are to achieve the strong coupling of $\Gamma=11$~GHz while keeping the $Q$s higher than $10^6$.
\begin{figure}[ht]
	\centering
	\includegraphics{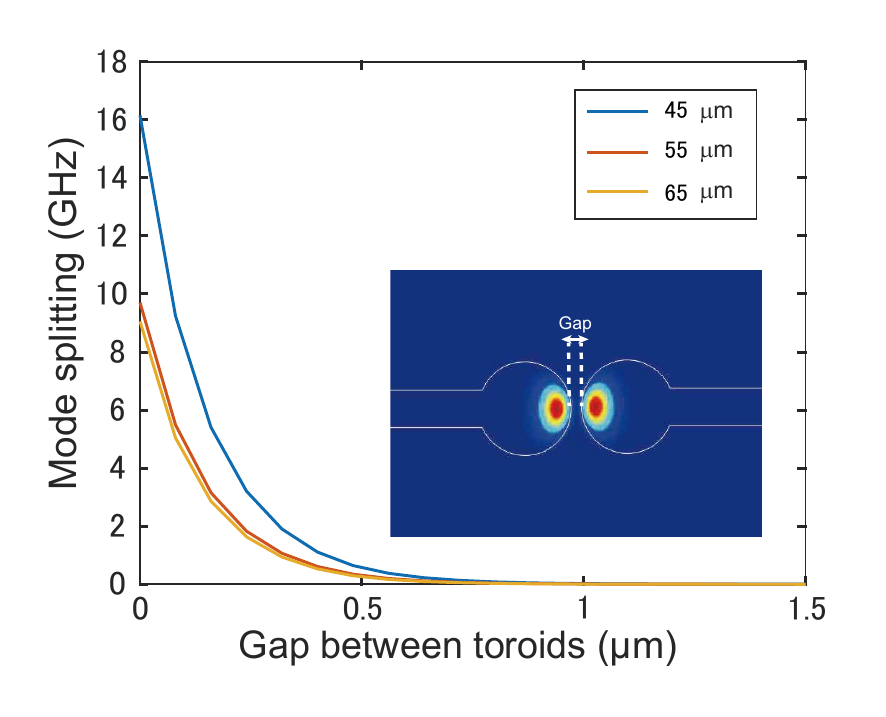}
	\caption{Simulation results showing supermode splitting as a function of the gap between silica toroids.}
	\label{fig:1}
\end{figure}


First, we calculate the splitting of the supermodes in coupled silica toroid microcavities and investigate the cavity geometries required to obtain the desired coupling strength.  A numerical study of the coupling coefficient between a tapered optical fiber and a microsphere has been reported \cite{HUMPHREY2007124}, and we extend this method to a coupled microcavity system. The coupling coefficient between cavity~1 (C1) and cavity~2 (C2) is obtained from the mode overlap between two microcavities and the phase matching condition. Taking these conditions into consideration, the coupling coefficient is given as,
\begin{equation}
\begin{split}
\tilde{\kappa}_{\mathrm{C1,C2}} &= \frac{\omega \epsilon_{0}}{4} (n^2-n^2_0) \times N_{\mathrm{C1}}N_{\mathrm{C2}} \iiint_{V_{\mathrm{C}}}((E_{\mathrm{C1}}(x,y,z)) \cdot \\
&\quad E_{\mathrm{C1}}(x,y,z)) \mathrm{e}^{i\Delta\beta z}dxdydz,
\label{eq:refname1}
\end{split}
\end{equation}
where $n$ and $n_{0}$ are the refractive indices of a silica toroid microcavity and air, $V_{C}$ is the cavity volume, $E_{C1}$ and $E_{C2}$ are the electrical fields of the two cavities, and $\Delta \beta$ is the propagation constant difference. Note that $N_{C1}$ and $N_{C2}$ are normalizing coefficients, which are given as,
\begin{equation}
\frac{1}{2} N^2_{\mathrm{C_1,C_2}} \sqrt{\frac{\epsilon_0}{\mu_0}} \iint n|\bm{E}|_{\mathrm{C_1,C_2}} dxdy =1,
\label{eq:refname2}
\end{equation}
Here we assume fundamental modes, and that the profile of the optical mode in each toroid is calculated by the finite element method (COMSOL Multiphysics) \cite{4230891}.  Although the mode profiles rigorously exhibit modulation when the toroids are strongly coupled, we assume that the mode profiles remain the same when we use Eq.~\ref{eq:refname1}.

The splitting of the supermodes $\Omega$ is given as,
\begin{equation}
\Omega \approx \frac{c}{2\pi nR}|\tilde{\kappa}_{\mathrm{C1,C2}}|^2,
\label{eq:refname3}
\end{equation}
where $c$ is the speed of light and $R$ is the diameter of the microcavity. We performed calculations for three different microcavity diameters $R$ of 45, 55 and 65 $\mu$m.  The calculation result is shown in Fig.~\ref{fig:1}, where the splitting of the supermodes is larger for a microcavity with a smaller diameter.  Importantly, this suggests that it is possible to obtain supermodes with a coupling strength exceeding 10~GHz when 55-$\mu$m-diameter microcavities are used.
\begin{figure}[htb]
	\centering
	\includegraphics{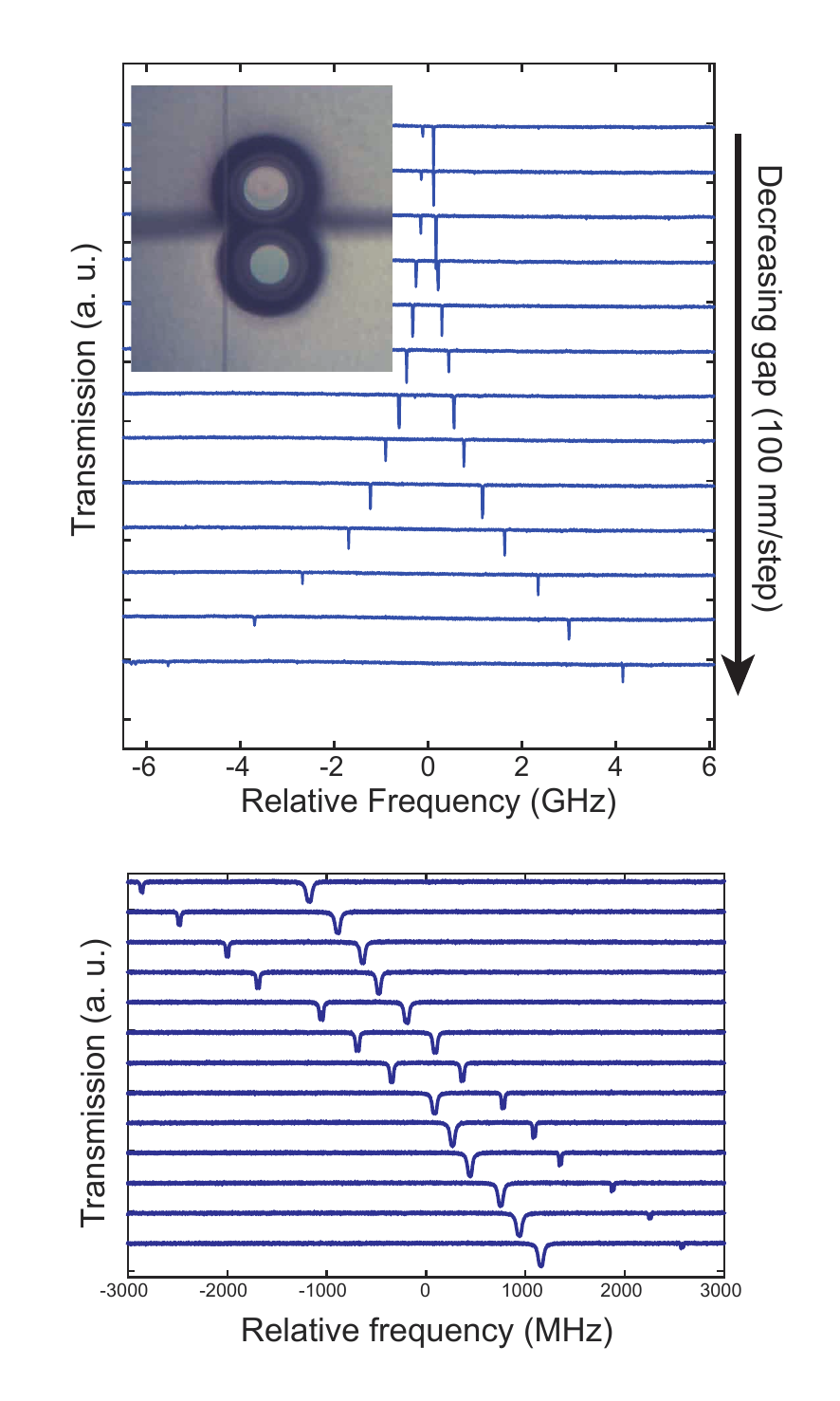}
	\caption{(a) The transmittance spectra exhibit different mode splitting widths when the gaps between two cavities are different. Inset: microscope image of coupled microcavities. (b) The transmittance spectra exhibit anti-crossing behavior when we change the temperature of C1, while the gap distance between the silica toroids is kept at about 1 $\mu$m.}
	\label{fig:2}
\end{figure}


We then fabricated two silica toroid microcavities with diameters of about 55~$\mu$m by using photolithography, XeF$_2$ vapor etching and CO$_2$ laser reflow.  They were fabricated on the edges of silicon chips by dicing the wafer after the photolithography \cite{Yoshiki2017}. Each silica toroid was placed on a 3-axis stage to enable high-precision alignment for adjusting the gap distance.  To conduct the experiment, it is necessary to select two modes that are close to each other in frequency.  It is also necessary to employ two optical modes with the same $Q$s to obtain Autler-Towns splitting \cite{PhysRev.100.703}, because Fano resonance \cite{RevModPhys.82.2257} appears when the $Q$s of the two optical modes are considerably different from each other.  In our experiments, we selected two optical modes with $Q$s of $1.5 \times 10^7$ and $1.3 \times 10^7$.

The inset in Fig.~\ref{fig:2}(a) shows the coupled cavity system that we used to measure the transmittance spectrum, where C1 is coupled to a tapered optical fiber.  The temperatures of the silica toroids were controlled separately by using a thermoelectric cooler, to make it possible to tune the resonant frequency via the thermo-optic (TO) effect.  We controlled the temperature of C2 so that we could tune its resonant frequency to match that of C1.  Although the resonant frequency of C1 moves slightly because the heat isolation between the two cavities is not perfect, we were able to tune the wavelength by about 100~pm, which is sufficient for our experiment.

After having adjusted resonances C1 and C2 to the same wavelength, we changed the gap distance between the two cavities.  The transmittance spectra are shown in Fig.~\ref{fig:2}(a), where the coupling becomes stronger as the gap decreases because the overlap between the modes in the two cavities becomes larger.  The spectrum shows that a very large mode splitting of $\Gamma > 10$~GHz is achieved.  It should be noted that this large splitting is achieved while maintaining a very narrow linewidth for the resonance modes $\Gamma_d$ (i.e. ultrahigh-$Q$) of the optical modes.  We achieved a record high $\Gamma / \Gamma_d \simeq 500$ is in our experiment, which was thanks to the ultrahigh-$Q$ of the toroid microcavity modes.

Figure 2 (b) shows the transmittance spectra for different temperatures of C1 when the gap distance between the silica toroids is about 1~$\mu$m. The result shows clear anti-crossing behavior, which proves that the cavity modes C1 and C2 are strongly coupled.

By adjusting the gap distance, we are able to obtain supermode splitting with frequency separation that is exactly the same as the SBS frequency shift.  As we discussed earlier, the advantage of this scheme is that we can demonstrate SBS lasing using small microcavities, and we do not require precise fabrication because the frequency separation is adjustable.

\begin{figure}[htb]
	\centering
	\includegraphics{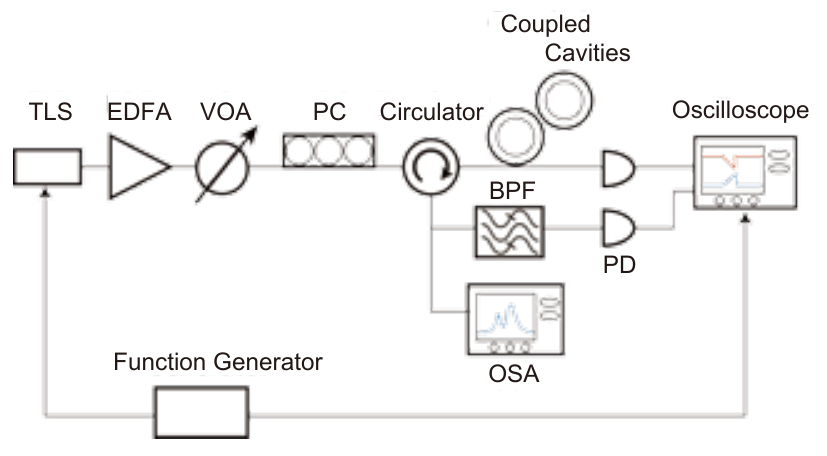}
	\caption{Experimental setup for Brillouin lasing. TLS: Tunable laser source. EDFA: Erbium-doped fiber amplifier. VOA: Variable optical attenuator. BPF: Band pass filter. PC: Polarization controller. PD: Photodetector. OSA: Optical spectrum analyzer.}
	\label{fig:3}
\end{figure}

Next, we performed experiments on the SBS in the coupled cavities. Figure~\ref{fig:3} shows the experimental setup. We employed a TLS to excite the pump mode. The pump light power was amplified with an EDFA and then controlled with a VOA.  The backward scattering light was sent to a PD connected to an oscilloscope and an OSA via an optical circulator. We used a narrow BPF to record only the backward SBS light. 

\begin{figure}[!h]
	\centering
	\includegraphics[width=7.5cm]{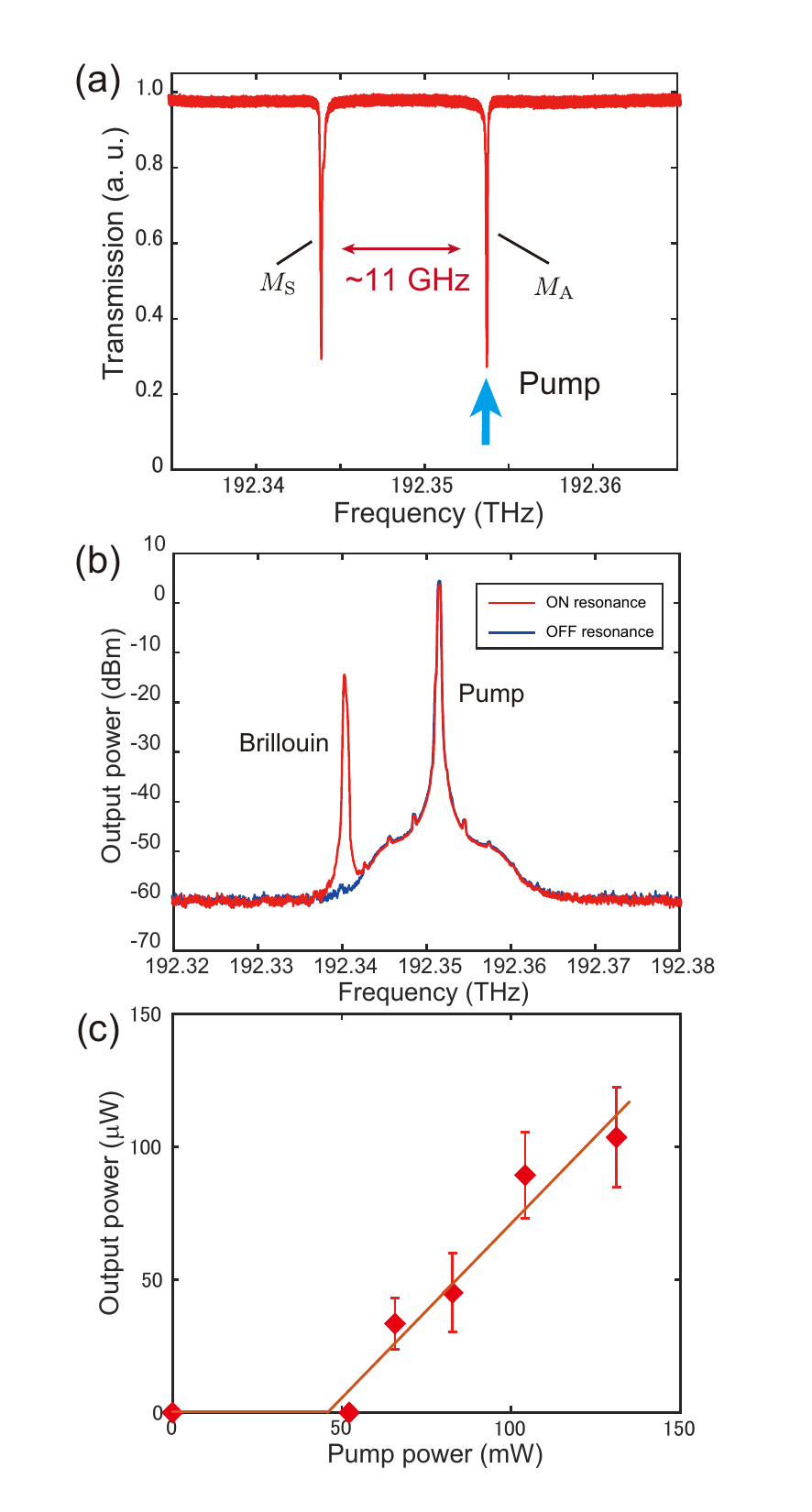}
	\caption{(a) Optical spectrum of the backscattered light. The pump signal is injected in high-frequency resonance ($M_{\mathrm{AS}}$: anti-symmetric mode), and the SBS light is generated in a low frequency mode ($M_{\mathrm{S}}$: symmetric mode). (b) Spectrum of SBS lasing behavior.  SBS light is observed when $\Gamma \simeq 11$~GHz (red line), while no lasing action is observed when $\Gamma\ne 11$~GHz. (c) SBS output power for different pump input powers.}
	\label{fig:4}
\end{figure}

Figure~\ref{fig:4}(a) shows the transmission spectrum of the supermode of the system we employed. The pump frequency is matched to a higher frequency of the anti-symmetric mode $M_{\mathrm{AS}}$ since the SBS light will be generated at a lower frequency. Using the function generator, the pump frequency is adjusted to obtain the maximum Brillouin output, because the resonant wavelength is red-shifted as the pump power increases. The symmetric mode in the supermodes had a resonant wavelength of 1558.57~nm and a $Q of \sim 2\times10^6$, and the anti-symmetric mode had a resonant wavelength of 1558.66~nm and a $Q of \sim2\times10^6$.  The splitting of the supermodes was adjusted to $\sim 11$~GHz to match the Brillouin frequency shift.  We observed a slight degradation in $Q$s when two toroids were placed close together, but the value remained sufficiently high to achieve SBS lasing.  We believe the $Q$ degradation in a coupled cavity system could be alleviated by selecting an appropriate mode pair as has been studied theoretically \cite{Boriskina:06,Boriskina:07}.

Figure~\ref{fig:4}(b) is the measured optical spectrum of the backscattered light.  The blue curve is the spectrum when the gap distance is not optimized.  We only observe one peak, which is the reflection of the pump. Since the splitting of the supermodes does not match the Brillouin frequency shift in silica, no SBS light is observed.  When we tune the gap distance and match the supermode splitting frequency to the Brillouin frequency shift of 11~GHz, we observe two peaks as shown by the red line in Fig.~\ref{fig:4}(b).  The peak on the lower frequency side is the SBS light, whose frequency shift from the pump agrees well with the Brillouin frequency shift.

Figure~\ref{fig:4}(c) shows the output power of the stimulated Brillouin scattering light as a function of the pump power.  The value is the recorded powers at the fiber input and output.   We repeated the experiment ten times, and the standard deviation is shown by error bars.  It shows clear threshold behavior, which indicates the presence of a stimulated process. 

We experimentally demonstrated SBS lasing in coupled microcavities, and obtained a small $P_{\mathrm{th}}$ of about 50~mW.  Although the value is small, it is higher than some of the previously reported values obtained using single WGM microcavities, because of the relatively low $Q$ we used in our experiment, which is not limited for any fundamental reasons.  The $P_{\mathrm{th}}$ value in a cavity is given as,
\begin{equation}
P_{\mathrm{th}} = \frac{\pi^2 n^2 V_{\mathrm{m}}}{B g_{\mathrm{B}} \lambda_{\mathrm{P}} \lambda_{\mathrm{B}} Q_{\mathrm{P}} Q_{\mathrm{B}}}.
\label{eq:4}
\end{equation}
where $n$ is the refractive index of silica, $V_{\mathrm{m}}$ is the effective mode volume, $B$ is the mode overlap, $g_{\mathrm{B}}$ is the Brillouin gain of silica, $\lambda_{\mathrm{P}}$, $\lambda_{\mathrm{B}}$, $Q_{\mathrm{P}}$, $ Q_{\mathrm{B}}$ are the wavelengths and quality factors of the pump and Brillouin lights, respectively \cite{Guo:15low}.  Equation~\ref{eq:4} shows that $P_{\mathrm{th}}$ is inversely proportional to the square of $Q$.  Since the coupled cavity system using a silica toroid can exhibit an ultrahigh $Q$ and a small mode volume, we should be able to reduce the threshold power significantly. And we can expect a value of $P_{\mathrm{th}}=5$~$\mu$W if we fabricate the sample carefully and employ cavities with $Q=10^8$.

Finally, we would like to discuss the advantage and differences when using a coupled cavity system rather than a single microcavity system to demonstrate SBS lasing.

First, the cascaded Stokes process is observed in a single microcavity when the pump power is increased. The odd Stokes lines are scattered backward for the pump, and the even Stokes lines are scattered forward. On the other hand, only the first Stokes line is observed in a coupled microcavity system. This is because the coupled cavity system only resonates with the first Stokes line unlike the single cavity system where equidistant resonances are present.

Secondly, the coupled microcavity system enables us to selectively enhance or suppress the Stokes line by changing the gap distance between microcavities. In the single cavity system, the optical mode frequency is fixed when the microcavity is fabricated.

Thirdly, as discussed above, we do not need to use a sub-mm-sized cavity to match the FSR with the Brillouin shift of the material.  Therefore, we can use cavities that are much smaller in size, which is advantageous for obtaining a smaller $P_{\mathrm{th}}$ as shown in Eq.~\ref{eq:4}.  Also as a result, we do not need to fabricate the cavity structure precisely.

In conclusion, we achieved the 11 GHz mode splitting of supermodes in coupled silica toroid microcavities. By employing small-diameter toroids, the $Q$ value of each supermode in the strong coupling regime was kept as high as 2 million. This large mode splitting enabled us to demonstrate the generation of stimulated Brillouin scattering using coupled cavity systems with a threshold of 50 mW by pumping with a 1550 nm-band tunable laser. When we compare these results with those obtained with other systems, the coupled toroid cavity system has the advantage that we can fabricate coupled silica toroids on-chip and do not require precise control of the cavity size. Our study may contribute to the miniaturization of microwave photonic devices.



%

\end{document}